\documentclass[11pt]{iopart}

\usepackage{amsfonts} 
\usepackage{amssymb}
\usepackage{setspace}

\usepackage[utf8,applemac]{inputenc}
\usepackage{tensor}
\usepackage{tikz}
\usepackage{graphicx}
\usepackage{iopams}
\usepackage{subfig}
\usepackage{wrapfig}

\eqnobysec

\newcommand{\bea}{\begin{eqnarray}}
\newcommand{\eea}{\end{eqnarray}}
\newcommand{\be}{\begin{equation}}
\newcommand{\ee}{\end{equation}}

\usepackage[unicode,breaklinks]{hyperref}

\begin{document}

\title[]{Are quantum corrections on horizon scale physically motivated?}


\date{\today}

\author{
Geoffrey Comp\`ere
}
\address{Université Libre de Bruxelles, CP 231, B-1050 Brussels, Belgium}

\eads{
\mailto{gcompere@ulb.ac.be}
}

\begin{abstract}
The aim of this short review is to give an overview to non-specialists of recent arguments from fundamental physics in favor and disfavor of quantum corrections to black hole horizons. I will mainly discuss the black hole information paradox, its possible resolutions and shortly address its relevance or irrelevance to astronomy. 
\end{abstract}



\tableofcontents
~
\clearpage
\newpage

\section{Some preliminary questions}

This article originates from a talk given in Athens, so starting with Socrates' method seems appropriate. 

\subsection*{1. What is a black hole?}

You might be sure to know the answer to this question and you might additionally think that you have the unique and correct answer. Yet, it has been argued that there is bias in answering this question depending whether you are a theoretical relativist, an electromagnetic wave astronomer, a gravitational wave astronomer or a theoretical quantum relativist \cite{Curiel:2019nna}. In order to be sure to have everybody on track, let me start by quickly reviewing the key concepts defining a black hole according to these 4 standpoints. 

For a theoretical relativist, a black hole is a classical metric with an event horizon, a singularity, it has thermodynamical properties, it is stable under small perturbations and it is the final state of collapse of heavy stars. That defines a Classical Black Hole. Note that in general relativity (GR), black holes have uniqueness properties but uniqueness is not anymore true in more general gravitational theories so this is not an essential feature of classical black holes. Also note that there are many examples of ``Black Hole Mimickers'' that are also classical objects conjectured to be the final state of collapse of stars in alternative theories of gravity but those don't have an event horizon nor a singularity so they are not classical black holes. 

From the standpoint of electromagnetism-based astronomy, black holes are essentially compact, electromagnetically black, massive objects which are compatible with the classical black hole of general relativity. For gravitational-wave astronomers, a black hole is first a source of gravitational radiation, again compatible with general relativity. 

Now, and this is becoming more relevant for the following, what is a Quantum Black Hole as defined by theoretical quantum relativists? Quantum black holes do not have an event horizon because they evaporate; they don't have a singularity because it ought to be regularized by quantum gravity effects; they should still have thermodynamical properties now enhanced by a microscopic statistical counting; they are not stable but only metastable due to Hawking radiation and finally, as we will discuss, some theoretical quantum relativists would also argue that they are not the end state of collapse as we usually define it classically. 

So, what do we actually really know about quantum black holes?

\subsection*{2. Do we know that black holes exist in Nature?}

As usual in physics, we never know for sure, so the answer is no but they are by far the best models to match the observations. If one uses the definition of a classical black hole, both the event horizon and the singularity are not observable neither by electromagnetic nor by gravitational wave observations. We only know the existence of compact, dark, massive objects emitting gravitational waves that are compatible with classical GR. Would you say otherwise?

\subsection*{3. Is Quantum Mechanics only relevant at microscopic scales?}

That's a very easy question. The answer is obviously no because long-range quantum coherent order exists. Examples include lasers, Bose-Einstein condensates, superfluids, superconductors, neutron stars, white dwarfs and many others. Ask yourself how a physicist would have answered this question 100 years ago at the advent of quantum mechanics.

\subsection*{3'. Do you think that Quantum Black Holes differ from Classical Black Holes only microscopically close to the singularity?}

That's the real question 3. Note that quantum gravity is still in its early phases; the theory is far from completion. There are two common answers to this question. The first is ``Yes, curvature is small away from the singularity so classical Einstein gravity is valid at the horizon.'' The alternative answer is ``No, a resolution of the Information Paradox requires long-range quantum order''. The correct answer can only come from theoretical quantum relativists. So I will spend some time reviewing arguments from that community in the following. 

\subsection*{4. If Quantum Black Holes are very different than classical black holes, can we observe (astronomically) these differences?}

That's an important question. The answer depends upon the nature of quantum black holes, which I need to address first. I will only make some comments on this at the end. 

\section{Foundations: what we know about Quantum Black Holes}

\subsection*{Thermodynamics: the Bekenstein-Hawking entropy}

Maybe one of the most beautiful formulae in the field of quantum gravity is the Bekenstein-Hawking entropy formula \cite{Bekenstein:1972tm,Hawking:1974sw}:
\bea
{S_{BH} = \frac{A}{4G} \frac{c^3}{\hbar} = \log W}.\label{BH}
\eea
Black holes have an entropy proportional to the area (!) of their classical horizon, with a specific factor of $1/4$ in front. The formula is universal for any matter minimally coupled to general relativity \cite{Iyer:1994ys}. In alternative theories of gravity, it is valid up to small corrections assuming that all couplings with respect to the Einstein action are small (in other words effective field theory is valid). The entropy formula obeys the second law of thermodynamics \cite{Hawking:1973uf}, which is a highly non-trivial property. Boltzmann told us that an entropy should count microscopic configurations, and indeed, this entropy formula can be reproduced for some non-astrophysical black holes in String Theory \cite{Strominger:1996sh,Berkowitz:2016jlq}, which is a second highly non-trivial property. No microscopic derivation of entropy has (yet) been performed for astrophysical black holes within String Theory. Yet, some alternative models of black holes explicitly violate this entropy formula at leading order, such as the Gravastar proposal \cite{Mazur:2001fv} or the proposal \cite{Chapline:2000en}. Such entropy formulae have not been shown to obey the second law of thermodynamics or be compatible with a microscopic counting. 

Finally note that the entropy formula points to a special role of the horizon in Quantum Gravity, and it also points to a notion of holography at the horizon, though I think that nobody really understands this. 

\subsection*{Thermodynamics: Hawking radiation}

Quantum black holes are not black. They radiate like\dots $\;$a black body. This is the famous 1974 Hawking computation \cite{Hawking:1974sw}. The black hole temperature is 
\bea
T = \frac{\hbar c^3}{8 \pi G M} \frac{2\lambda}{1+\lambda},\qquad \lambda = \sqrt{1-\frac{a^2}{M^2}}. \label{T}
\eea
You have probably mostly seen the first part of the formula which is valid for the Schwarzschild black hole. The second multiplicative Kerr correcting factor is 1 for Schwarzschild ($a=0$) and decreases to 0 for extreme Kerr ($a=M$).
This computation is done for a black hole in the vacuum. Astrophysical black holes are not in the vacuum. They are surrounded by the cosmic microwave background (CMB) at $2.7 K$. Plugging the numbers, all known solar mass or supermassive black holes have a Hawking temperature way lower than the CMB temperature and therefore absorb radiation. We would need black holes to have a ``very small'' mass $M < 10^{22}$ kg to start to emit radiation, but no such black holes have ever been observed. It has been conjectured that such black holes could have been formed from quantum fluctuations in the early universe \cite{Carr:1974nx}. 

Black holes in the vacuum radiate and therefore evaporate. The timescale for evaporation is
\bea
t_{evap} \sim \frac{G^2 M^3}{\hbar c^4} \sim 10^9 \frac{M^3}{(10^{11} kg)^3} yrs
\eea
Tiny black holes of mass $M < 10^{11} kg$ would evaporate within the current Universe time. Again, no one has seen those. Astrophysical black holes will evaporate eventually, but in a really long time. Black hole evaporation is therefore mainly discussed as a thought experiment.

\subsection*{Penrose diagram}

\begin{figure}[h!]
\subfloat[Semi-classical view]{
\includegraphics[width=0.30\textwidth]{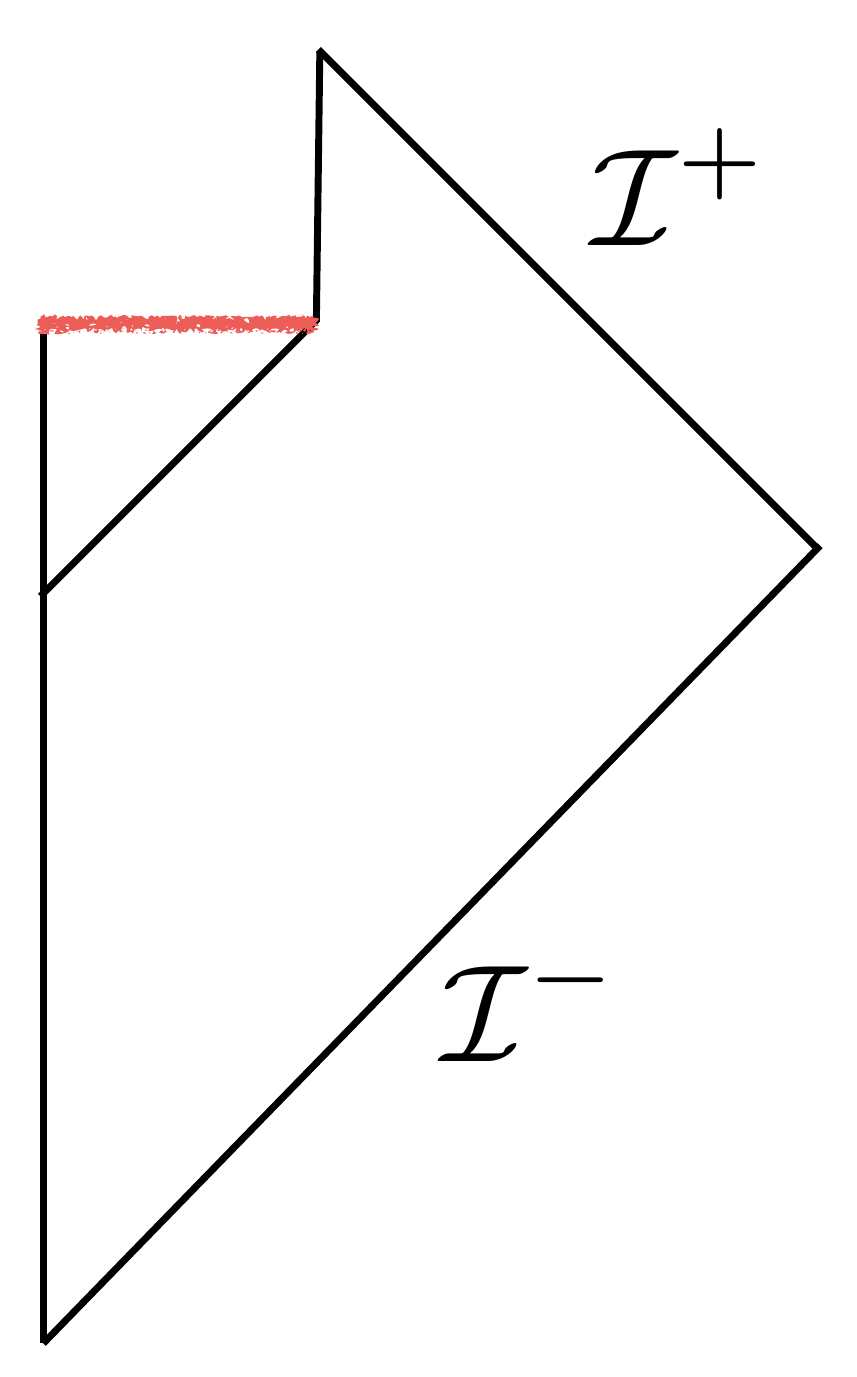}
}\hfill
\subfloat[Agnostic view]{
\includegraphics[width=0.36\textwidth]{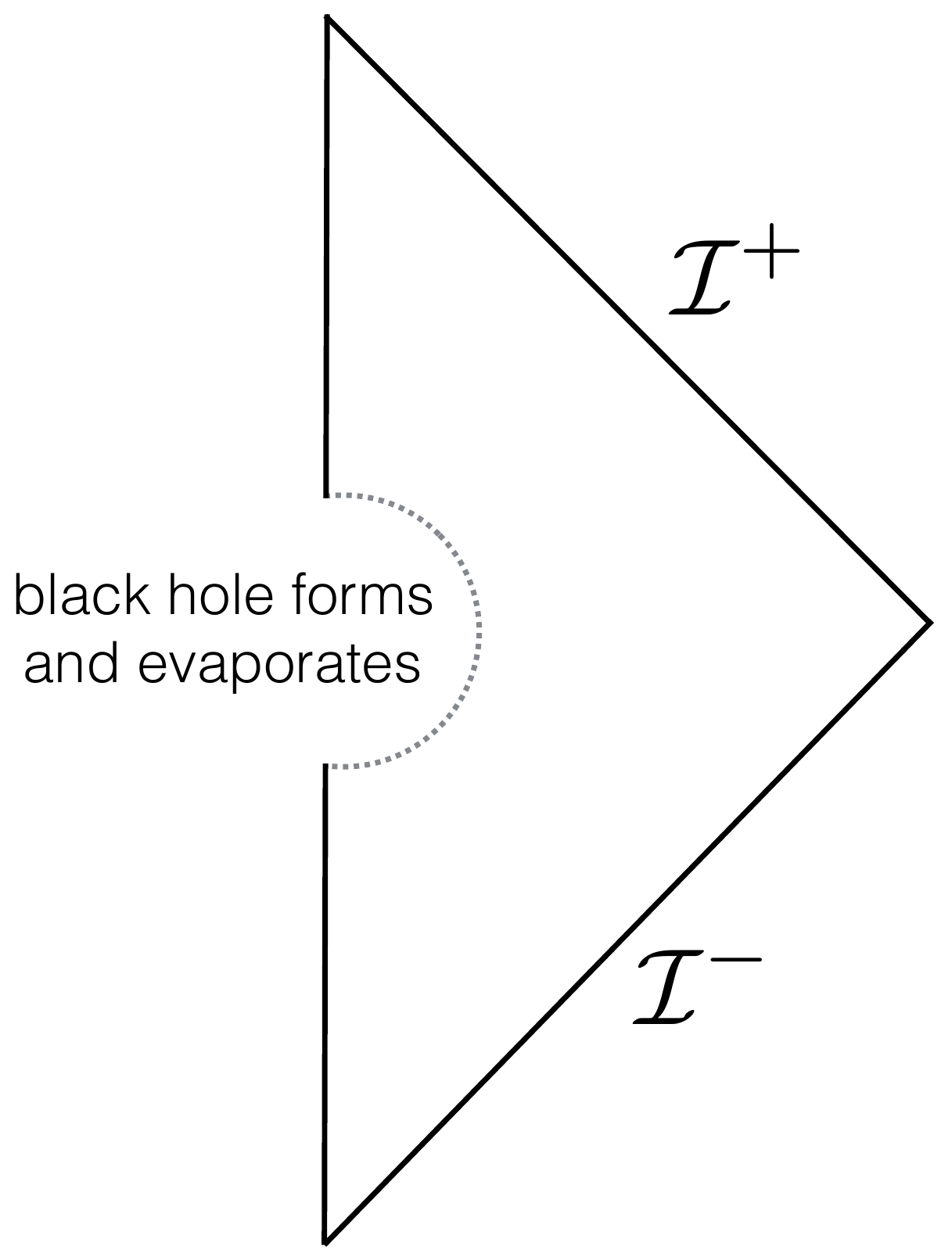}
}
\caption{Penrose diagrams of an evaporating black hole.}\label{fig:Penrose}
\end{figure}

The causal structure of a spacetime can be depicted by a Penrose diagram. There are two more-or-less standard pictures that depict a quantum black hole, which I will call the semi-classical view, and, the agnostic view, see Fig. \ref{fig:Penrose}. 

The semi-classical view contains the singularity and the horizon. Both of these are classical concepts that require a change in the quantum theory. In this diagram, the horizon is cut after the time of evaporation, leaving afterwards a portion of Minkowski space (after assuming no remnants, see further on). If one draws a timelike curve around the horizon (e.g. one  takes an observer sitting at finite distance away from the horizon) and replaces the inside spacetime with an ``ignorance patch'', one gets the second agnostic picture. Note that at this point there can be a baby universe with another asymptotic boundary inside the ignorance patch (obtained by resolving the singularity in the semi-classical picture). The picture only describes the exterior universe to the black hole. We will go back to this in a moment.

\subsection*{Some general statements about quantum gravity}

To move on, I find useful to give a couple of elements on quantum gravity. First, GR is not a quantum theory in a standard sense because it is perturbatively non-renormalizable \cite{Goroff:1985th}. Instead, GR is an effective theory at low energies. It also has to do a lot with thermodynamics, see Jacobson's derivation of GR from the Raychaudhuri's equation, the Bekenstein-Hawking entropy formula, and the first law of thermodynamics \cite{Jacobson:1995ab}. 

Also, spacetime is generally thought to be emergent after averaging over highly non-local quantum gravitational degrees of freedom. The non-locality of the vacuum is already present in quantum field theory, see e.g. the Reeh-Schlieder theorem \cite{Witten:2018lha}. An explicit realization of the emergence of spacetime is the AdS/CFT correspondence \cite{Maldacena:1997re}, which is a definition of quantum gravity in specific settings. 

\section{The Paradox}

\subsection*{Hawking's 1974 Information Paradox}

Hawking's reasoning \cite{Hawking:1974sw} can be loosely stated as follows. Around the horizon of a black hole, pairs are created in an entangled state
\bea
\psi_{pair} = \frac{1}{\sqrt{2}}(e_{in}^- e_{out}^+ + e^{+}_{in} e^-_{out}). 
\eea
Let's say that either an electron gets in and a positron comes out or a positron gets in and an electron comes out.  In fact, neither happen, the state is a perfect superposition of the two. There is a net zero energy and all particles coming out have positive energy because they are long-lived and observable. The ingoing particle therefore has negative energy. This is the result of a computation of quantum fields in the Schwarzschild or the Kerr geometry. When one particle is absorbed by the black hole, the mass of the black hole is lowered (because a negative energy particle is absorbed) and the entanglement entropy of the outgoing radiation is increased by $\ln 2$ (because of the perfect  entanglement of the Hawking pair). No information of the black hole is emitted. 

The process goes on, until all the black hole has evaporated, leaving a radiation entangled with nothing, that is, a density matrix. This violates the unitary evolution of Quantum Mechanics of the exterior spacetime. 

Hawking's computation is robust. Only small flaws have been found that do not change the conclusion \cite{Wald:1999vt}\footnote{In particular, Hawking assumes that the collapsing matter is in the vacuum state. Assuming instead an excited state leads to additional stimulated emission. However, it does not change the late time black body spectrum and the paradox remains unchanged \cite{Wald:1976ka,Bekenstein:1977mv,Panangaden:1977pc}. For an alternative view, see \cite{Lochan:2016nbs} and the review \cite{Chakraborty:2017pmn}.}. There are also no-go theorems stating that small modifications cannot alter the result \cite{Mathur:2009hf}.

\subsection*{Modern version of the Information Paradox}
\label{sec:mod}

A modern version of the paradox consists in stating a list of hypotheses that are sufficient to arrive to a logical contradiction. Therefore, one (of several) of the hypotheses has to be wrong. The Information Paradox is that one of the following claims is wrong (see the reviews \cite{Mathur:2009hf,Harlow:2014yka,Polchinski:2016hrw,Marolf:2017jkr}):
\begin{itemize}
\item Evolution is unitary as defined by a distant observer
\item Remnants and Baby universes do not exist
\item The outside vacuum does not encode the black hole microstate
\item Effective Field Theory applies at large distances (as compared to Planck/string scale) outside the horizon
\item Quantum gravity effects are exponentially suppressed at microscopic distances around the horizon.
\end{itemize}
A remnant is an hypothetical arbitrarily high entropic state of Planckian mass. A baby universe is an hypothetical asymptotically flat spacetime that observers falling inside the black hole can reach. The other sentences are self-explicatory. If one of the first 5 claims is wrong, it solves the paradox. The above list is usually assumed to be complete. A logical possibility is that a tricky loophole  that nobody was able to pin down hides somewhere, but I shall not consider this possibility further. 

The original take of Hawking on this is that information is forever lost into the black hole. There might be unitarity in a larger sense, taking into account part of the state that fell into the black hole, but there is no unitarity as seen by the exterior observer alone. It means that given the data  of a complete Cauchy slice before gravitational collapse into a black hole, the outside observer will not be able to construct the final state of evaporation of the black hole. This leads to a loss of predictability. This take is still taken today by part of the (quantum) relativity community, see e.g. \cite{Unruh:2017uaw}. 
 
\subsection*{Claims from AdS/CFT}

The AdS/CFT correspondence \cite{Maldacena:1997re,Witten:1998qj} is a definition of quantum gravity in terms of a conformal field theory in specific settings in string theory. It allows to answer questions on quantum gravity that cannot be answered otherwise, i.e. without a UV complete quantum gravity theory. The main limitation of the AdS/CFT correspondence is that it only applies to gravity embedded in string theory (in particular in 10 dimensions) and it only applies to asymptotically anti-de Sitter spacetimes. 

\begin{wrapfigure}{r}{0.30\textwidth}
\includegraphics[width=0.35\textwidth]{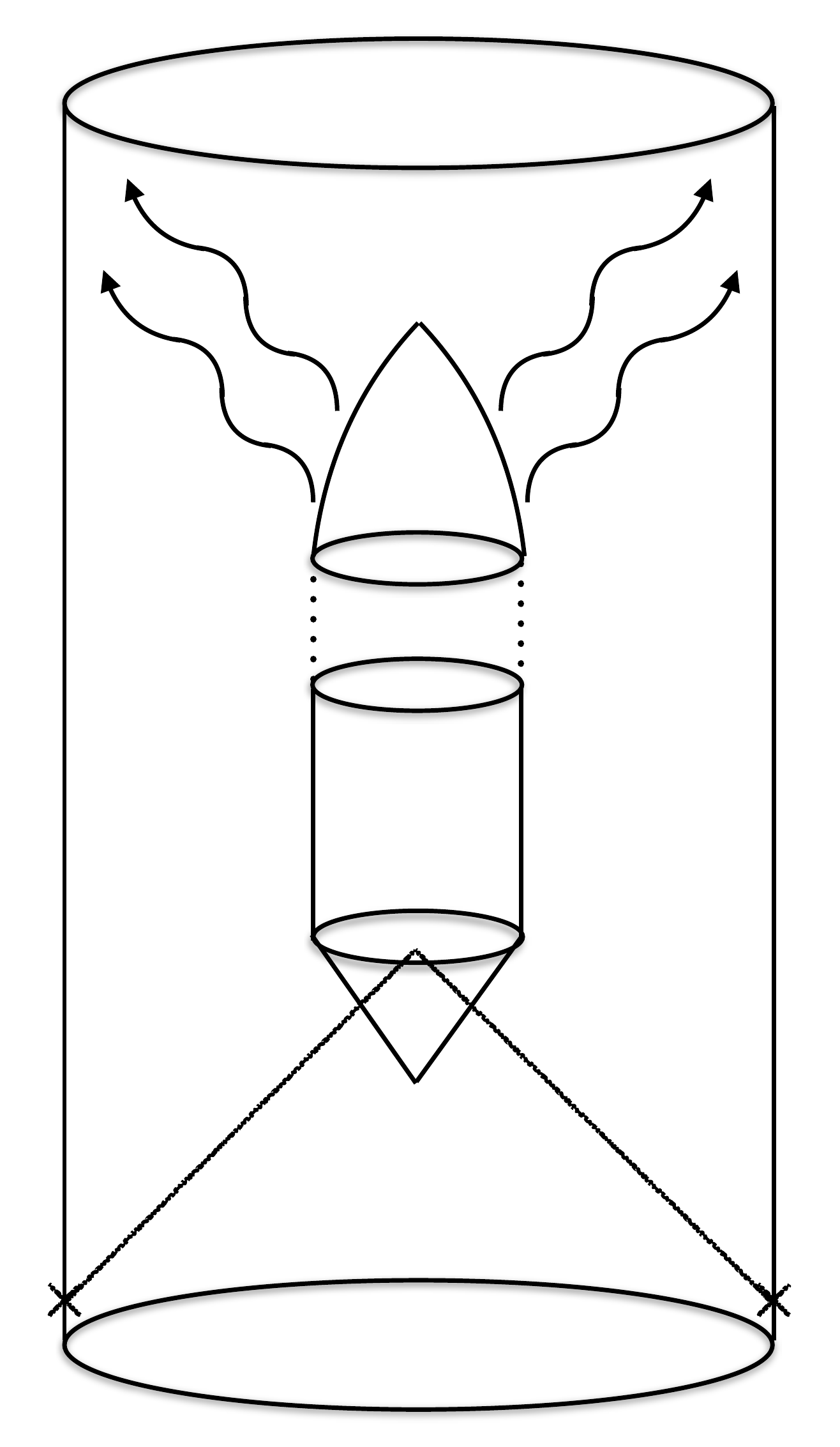}
\caption{Evaporating black hole in AdS.}
\label{AdSBH}
\end{wrapfigure}

In AdS/CFT, one can argue that (i) evolution is unitary as seen by a distant observer and (ii) Remnants or Baby universes do not exist. The argument is the following. We consider the creation of a black hole by incoming radiation. We consider a small black hole with size less than the AdS radius. The black hole evaporates in time $\sim M^3$. The creation of the black hole is in one-to-one correspondence in the CFT to a certain action of operators on the vacuum state. The evaporation process is in one-to-one correspondence with the unitary time evolution of the CFT. Since the initial state before the black hole was formed evolves unitarily to a state of pure radiation after the black hole has evaporated, the Hawking radiation process is unitary. In particular, there is no loss of information in a baby universe ``inside the black hole horizon''. Since there is no low energy, highly entropic state in a CFT, there is no corresponding remnant in the gravity theory. 

These arguments are established using an AdS box around the black hole, and in the framework of string theory with additional fields and dimensions. The claim is then that the results also extend to 4d asymptotically flat black holes in any consistent  UV completion of Einstein gravity. This excludes Hawking's take, and he himself subsequently revised his position \cite{Hawking:2005kf}. Yet, not all relativists agree with these AdS/CFT arguments \cite{Unruh:2017uaw}.

Another argument against the baby universe scenario and for exterior unitarity, but more general than the AdS/CFT argument, is the following \cite{Marolf:2008mf}. In gravitation, the Hamiltonian is a boundary term. Therefore, gravity is holographic in the following sense: by knowing all energy states at spatial infinity, one knows all possible spacetime states. This excludes baby universes that could not be described by the boundary Hamiltonian. Moreover, by unitary evolution at the boundary, all spacetime states need to evolve unitarily in the bulk of spacetime. The AdS/CFT correspondence is just one example of holographic theory where these expectations are realized.

\subsection*{The outside vacuum does not encode the black hole microstate}

In quantum field theory, there is a unique Lorentz-invariant vacuum defined with respect to a globally defined timelike Killing vector. In curved spacetime, there is no unique definition of time foliation, and therefore, no unique definition of positive frequency modes and vacuum. Asymptotically flat spacetimes asymptote to Minkowski and, therefore, a quantum field theory vacuum can be defined in the asymptotic region. Recently, it has been observed that in quantum gravity, the asymptotically flat vacuum is not unique \cite{Strominger:2013jfa}. This has led to an interesting infrared triangle relationship between soft theorems, the displacement memory effect and supertranslation symmetries \cite{Strominger:2017zoo}. 

The non-uniqueness of the vacuum has suggested a new way out of the paradox using ``soft hair'', i.e. nearly zero energy eigenstates \cite{Hawking:2016msc}. The proposal is roughly that soft hair encode the black hole microstates and allow to formulate a unitary evolution of black hole evaporation\footnote{For earlier work on vacuum degeneracy related to the black hole paradox, see \cite{Hossenfelder:2012mr,Hossenfelder:2014jha}.}. Nearly zero energy eigenstates have nearly infinite wavelength and therefore extend to the asymptotically flat region. Hawking radiation is not influenced by soft hair \cite{Javadinazhed:2018mle,Compere:2019rof}, which is consistent with the latest conjecture \cite{Strominger:2017aeh}.  Now, the proposed resolution is problematic in several respects. Soft modes factor out from the S-matrix \cite{Bousso:2017dny} though correlation might remain between finite energy states and soft radiation \cite{Carney:2018ygh}. There is no indication that there is enough information in soft hair to encode the black hole entropy (or worse the one of $N \rightarrow \infty$ distinct black holes that share the same asymptotically flat region). Finally, the proposed resolution is using specific features of asymptotically flat spacetimes while the paradox can also be formulated in $AdS$. 

\section{Sharpening the Paradox}

\subsection*{What we are left with}

After discarding the first three items listed in Section \ref{sec:mod}, one is left with the conclusion that one of the following hypotheses is wrong: 
\begin{itemize}
\item Effective Field Theory applies at large distances (compared to Planck/string scale) outside the horizon
\item Quantum gravity effects are exponentially suppressed at microscopic distances around the horizon.
\end{itemize}
This allows to understand the answer to Question 3' of several prominent members of the quantum theoretical relativity community. The claim is that a resolution of the Information Paradox requires long-range quantum order, extending at least to the would-be classical horizon. 

\subsection*{The Tunnelling Argument}

Let us discuss the arguments for long-range quantum order based on Quantum Gravity Euclidean path integrals. This path integral is not well-defined because Einstein gravity is perturbatively non-renormalizable, but there exists motivated prescriptions in several cases to remove the UV divergences. Using such prescriptions and a particular boundary condition for metric fluctuations at large distances, the Euclidean Schwarzschild black hole has action \cite{Gibbons:1976ue}
\bea
I_E = 4 \pi M^2. 
\eea
The Einstein-Hilbert action vanishes because the Ricci is zero, but there is a Gibbons-Hawking boundary term that brings this contribution. Therefore, the probability to tunnel to the Schwarzschild black hole at fixed temperature is \cite{Gross:1982cv}
\bea
e^{-I_E} = e^{-4 \pi M^2}=e^{-\frac{1}{2} \frac{M}{T}}  
\eea
where the mass was converted into temperature using the Hawking temperature formula (\ref{T}) for the non-rotating black hole. One can generalize this computation to more general black holes (e.g. the Kerr black hole or higher dimensional black holes). It turns out that in each case the Euclidean action is equal to the Gibbs potential $\Phi$ of the black hole divided by the temperature \cite{Gibbons:2004ai}. For a non-rotating black hole\footnote{Naively, the Euclidean action for the Schwarzschild black hole equals the Bekenstein-Hawking entropy, $S_{BH} = \frac{4 \pi (2M)^2}{4}=4 \pi M^2$ after using the area formula $A=4\pi R^2$ with Schwarzschild radius $R=2M$. However, this is not true in general (e.g. a higher dimensional black hole or a Kerr black hole). It would be therefore incorrect to state that $I_E = S_{BH}$ as a thermodynamical equality.}, 
\bea
I_E = \frac{\Phi}{T}=  \frac{M- T S_{BH}}{T}
\eea
The probability to tunnel to the Schwarzschild black hole can therefore be written in the suggestive form
\bea
e^{-I_E} = e^{-\frac{M}{T}} e^{S_{BH}}.
\eea
Since there are $e^{S_{BH}}$ microstates, that equation suggests that there is a tunneling probability for each individual microstate given by the Boltzmann factor $e^{-\frac{M}{T}}$. 

Now, it is not clear that the Boltzmann factor $e^{-\frac{M}{T}}$ is the relevant distribution for quantum gravity microstates. After all, quantum mechanical systems have distinct probability distributions (the Fermi-Dirac and Bose-Einstein distributions). Instead, arguments were given that the probability to tunnel to a single black hole microstate is given by \cite{Mathur:2008kg,Kraus:2015zda,Bena:2015dpt}
\bea
e^{-S_{BH}}.
\eea
Since the number of distinct black hole microstates is given by the Bekenstein-Hawking formula (\ref{BH}), $e^{S_{BH}}$, the probability of quantum tunneling to a quantum gravity microstate during collapse, instead of a black hole, would then be of the order of $e^{-S_{BH}} e^{S_{BH}} = 1$. This could only be possible due to a long-range quantum gravity order that prevents the formation of a black hole. In this scenario, the natural question is whether one can distinguish a microstate from a black hole, but that question is unconclusive so far \cite{Guo:2017jmi,Cardoso:2019rvt}.

\subsection*{The Page curve}

\begin{figure}[!h]\vspace{-0.3cm}
\subfloat[Definition of early (R) and late (BH) radiation with respect to an asymptotic observer's retarded time. Tracing out the late (BH) radiation gives the entanglement entropy depicted in (b). Credit: \cite{Harlow:2014yka}. ]{
\includegraphics[width=0.30\textwidth]{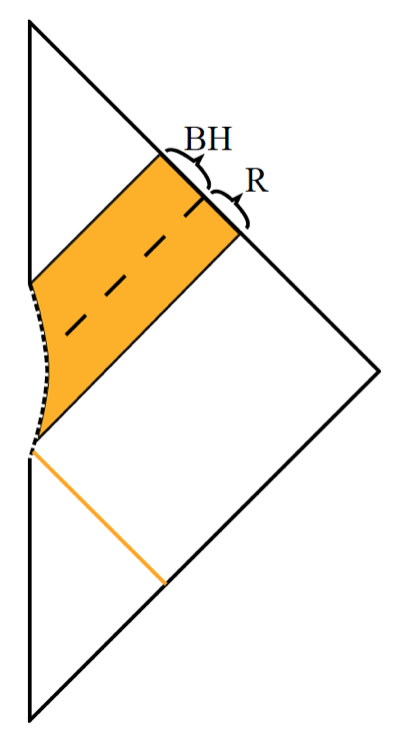}
\label{fig:Harlow1}
}\hfill
\subfloat[Two expectations for the shape of the entanglement entropy curve. Hawking's curve is linear. Page's curve reaches zero at evaporation time.]{\vspace{-1.3cm}
\includegraphics[width=0.52\textwidth]{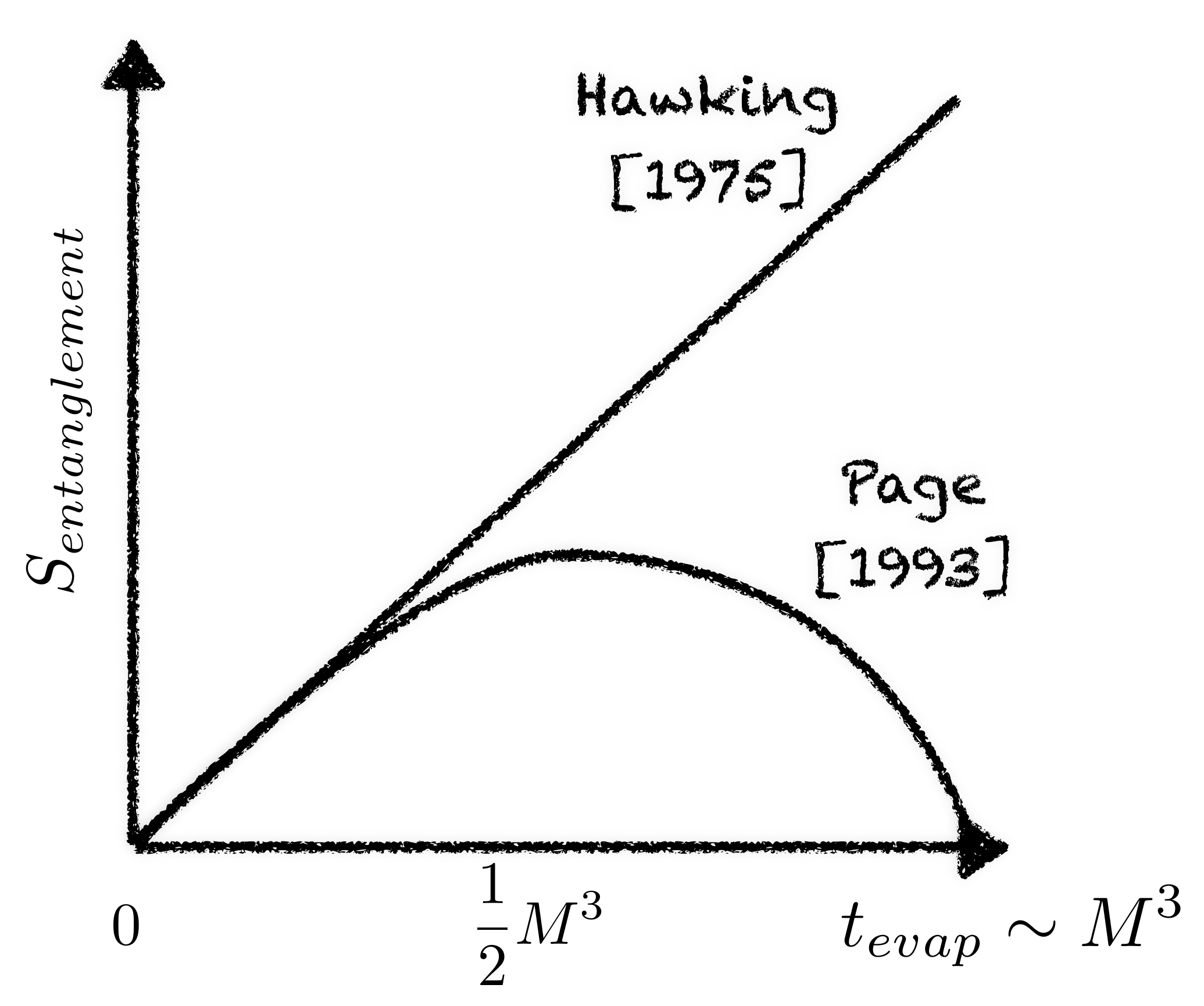} 
\label{fig:PageCurve}
}\vspace{-0.3cm}
\caption{}
\end{figure}

Quantum black holes can be described using tools of quantum information theory. A lot of progress has been obtained in this area recently (see e.g. \cite{Harlow:2014yka}). One of the early results on the quantum information theory of black holes is the so-called Page curve \cite{Page:1993wv}. 
Let us imagine a black hole produced in the vacuum by an infalling null shell. Such a black hole Hawking radiates. At a given retarded time as measured by a far observer, the Hawking radiation can be separated into two sets. The earlier set that already radiated to future null infinity and the future set that will be emitted later and that encodes the state of the black hole at that retarded time. See Figure \ref{fig:Harlow1} taken from Harlow's lectures \cite{Harlow:2014yka}. 

According to Hawking's result \cite{Hawking:1974sw}, each time a Hawking pair is produced, the entanglement entropy of the later radiation after tracing out the earlier radiation is increased by $\ln(2)$. This leads to the monotonic increase of entanglement entropy all the way to the final evaporation time $\sim M^3$. Instead, assuming a unitary radiation process as seen by an exterior observer, the entanglement entropy has to start dropping at least half-way $\sim \frac{1}{2}M^3$ and reach 0 at evaporation time. The resulting curve of entanglement is called the Page curve \cite{Page:1993wv}, see Figure \ref{fig:PageCurve}.

\subsection*{The AMPS Paradox}

Quite recently, a reformulation of the Hawking paradox was formulated by Almheiri, Marolf, Polchinski and Sully \cite{Almheiri:2012rt} and led to a flurry of activity. More than a reformulation, it led to disprove some earlier conjectured resolutions of the paradox such as the proposed ``Black Hole Complementarity'' \cite{Susskind:1993if}, see comments in \cite{Almheiri:2013hfa,Susskind:2014moa}. 
 
The argument can be summarized as follows. Consider an outgoing Hawking mode emitted after Page time. Unitarity of the outside radiation implies that that mode is entanged with the early radiation. Let us also assume that infalling observers can use the Kerr metric with exponentially suppressed quantum corrections. Then the outgoing Hawking mode is highly entangled with an ingoing mode behind the horizon. This collection of mutual entanglements exceeds the limit allowed by quantum mechanics. More precisely, the strong subadditivity theorem of entanglement entropy is violated. This is a contradiction. 

Again, we are forced to drop one of the following hypotheses: 
\begin{itemize}
\item ``EFT'': Effective Field Theory applies at large distances (compared to Planck/string scale) outside the horizon;
\item ``No drama'': Quantum gravity effects are exponentially suppressed at microscopic distances around the horizon, or in other words, infalling observers encounter nothing unusual at the horizon; 
\item Quantum Mechanics holds; 
\end{itemize}
or, of course, another hidden hypothesis.

\section{Speculative proposals for resolving the Paradox}

At this point in the reasoning, there are 3 main options to solve the paradox, which consist in dropping either one of the three items above:  ''No Drama'', ``EFT'' or Quantum Mechanics. There is no consensus on the resolution, so all of them are speculative. Here is a partial list of proposals. I don't mean at all to be exhaustive and I'm sorry not to have included many other relevant options; 

\subsection*{Prop 1. Violation of the Equivalence Principle}

\subsubsection*{Prop 1.a) Firewall}

\begin{wrapfigure}{r}{0.30\textwidth}\vspace{-0.7cm}
\includegraphics[width=0.20\textwidth]{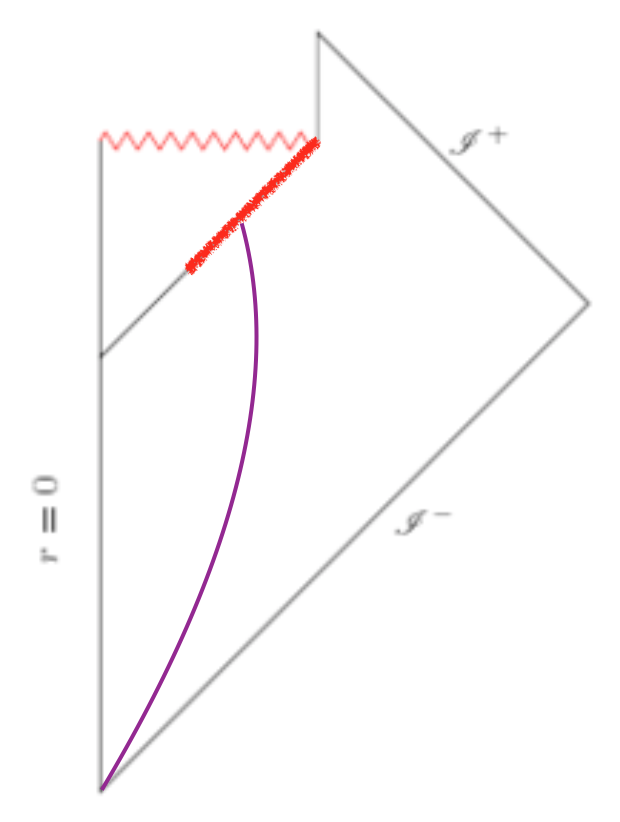}
\caption{Firewall.}\vspace{-1cm}
\label{Firewall}
\end{wrapfigure}
The take of AMPS \cite{Almheiri:2012rt,Almheiri:2013hfa} is that there is drama for the ingoing observer. There is a ``firewall'' that forms after some time in the black hole evaporation process (either at Page time or earlier) that explicitly violates the equivalence principle and leads to strong quantum effects acting on the ingoing observer. However, there is no explicit model of a firewall in quantum gravity. For a proposal for semi-classical firewall, see \cite{Kaplan:2018dqx}. Critics call this proposal \dots dramatic.

\subsubsection*{Prop 1.b) Fuzzball}

\begin{wrapfigure}{r}{0.30\textwidth}\vspace{-0.9cm}
\hspace{0.5cm}\includegraphics[width=0.20\textwidth]{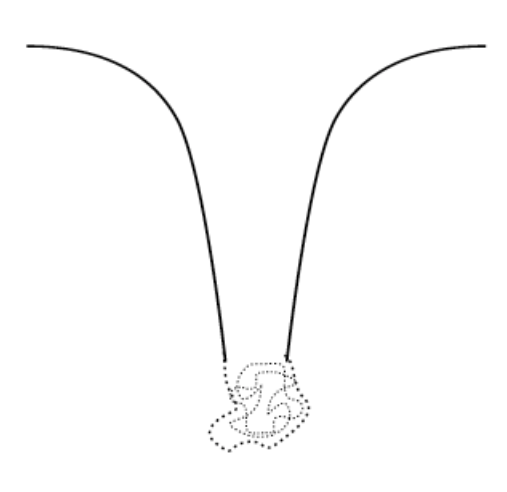}
\caption{Fuzzball. Credit: \cite{Mathur:2005zp}.}
\label{Fuzzball}\vspace{-0.5cm}
\end{wrapfigure}
The fuzzball resolution is an earlier idea originating in some explicit AdS/CFT models where it is explicitly realized \cite{Lunin:2001jy}. One starts from the Tunnelling Argument reviewed earlier: one conjectures that there are $e^{S_{BH}}$ quantum gravity coherent states that describe the microstates of the black hole. One further conjectures that the microstates are described by a semi-classical non-GR higher dimensional theory of supergravity also including classical features of string theory. For more details, see e.g. \cite{Mathur:2005zp,Skenderis:2008qn,Bena:2013dka}. Critics say that there has never been a proof that there are enough such states to describe the Kerr black hole. Moreover, the hypothesis of classicality can be debated since the fine-grained structure required to detail each individual microstate might be too small to be described by classical objects. 

\subsection*{Prop 2. Violation of the Effective Field Theory far from the black hole}

\subsubsection*{Prop 2.a) Entanglement implies Wormholes}

\begin{wrapfigure}{r}{0.30\textwidth}\vspace{-15pt}
\includegraphics[width=0.30\textwidth]{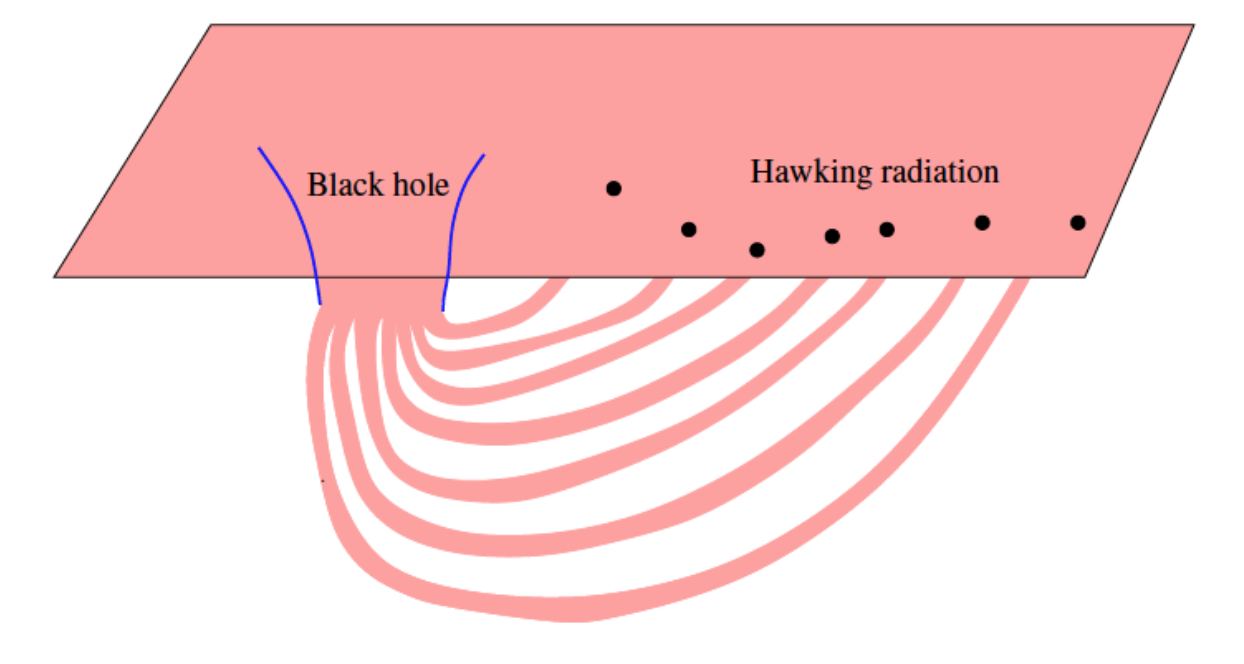}
\caption{Entanglement implies Wormholes. Credit: \cite{Maldacena:2013xja}.}
\label{EW}
\end{wrapfigure}

The Einstein-Rosen bridge between the two asymptotically AdS regions of an eternal AdS black hole is a non-traversable wormhole. It has been recently shown in the context of the AdS/CFT correspondence that this geometry can be understood in terms of entangled states of two distinct conformal field theories. This suggests a more general connection between wormholes and entanglement \cite{Maldacena:2013xja}. A conjecture is then that the high entanglement between a black hole beyond Page time and the Hawking radiation leads to a wormhole-like structure that remains to be understood.  There is however no model of such conjectured wormholes. It has also been argued that this idea by itself is not sufficient to solve the paradox  \cite{Susskind:2014moa}. 

\subsubsection*{Prop 2.b) New effective non-local interactions}

There are also attempts to set a framework of fundamental principles to constraint the type of non-local interactions needed to solve the paradox \cite{Giddings:2006sj,Giddings:2017mym}. A toy model along these lines has been proposed \cite{Osuga:2016htn}, see also \cite{Yosifov:2018yft}. Such scenarii have been argued to lead to $O(1)$ corrections of any signal propagating around the horizon scale \cite{Giddings:2017jts}. Critics have pointed that there is a lack of testable models including in the AdS/CFT correspondence, and that it is not clear whether such non-local interactions can solve the paradox at first place because of a lack of detailed balance \cite{Marolf:2017jkr,Almheiri:2013hfa}. A semi-classical mechanism that might obey detailed balance was recently proposed \cite{Emelyanov:2017dcd}.

\subsection*{Prop 3. Modifications of Quantum Mechanics}

\subsubsection*{Prop 3.a) Non-locality from state-dependence}

Motivated from the AdS/CFT correspondence, another form of non-locality was proposed based on the state-dependence of observables \cite{Papadodimas:2012aq,Papadodimas:2013wnh}. It was however shown that such a formulation breaks the Born rule of quantum mechanics \cite{Harlow:2014yoa,Marolf:2015dia} so the proposal is not final.

\subsubsection*{Prop 3.b) Other modifications}

Modifying quantum mechanics in a consistent manner is known to be a very hard task. Several modifications of quantum mechanics have been proposed (see e.g. \cite{Hartle:2018yul}) but it is not clear whether they lead to a resolution of the paradox. 

\section{Final comments: quantum gravity physics with gravitational wave detectors?}

According to the recent arguments advanced by the AdS/CFT correspondence, the process of black hole evaporation in the vacuum is unitary as seen by an exterior observer. In order to solve the Information Paradox, one is then led to the conclusion that either quantum black holes admit some new structure at the horizon scale, or, that effective field theory breaks down due to a new form of non-local physics, or, that quantum mechanics needs to be corrected. Today, there is no well-accepted model of a quantum black hole by quantum relativists. Known astrophysical black holes are very young relative to the timescale of black hole evaporation, which makes the role of such quantum considerations unclear for astrophysical purposes.

Echoes in post-merger gravitational wave observations (as well as absorption in the inspiral phase of binary mergers \cite{Cardoso:2019rvt}) have been proposed as potential signatures of new quantum horizon physics or of new exotic compact objects \cite{Cardoso:2016oxy}. From the standpoint of fundamental physics, there is no first principle computation so far  in quantum gravity showing that the absorption rate of a quantum black hole will be away from unity for astrophysical black holes, though it is not excluded. Other potential observational signatures include changes in the tidal deformability and a distinct multipolar structure than the Kerr black hole \cite{Cardoso:2019rvt}. Such signatures are being investigated from available gravitational wave data, see e.g. \cite{Abedi:2016hgu}.

An important problem to be faced by theorists that attempt to derive signatures of quantum black holes is degeneracy: even if there is a non-GR signal observed at LIGO/Virgo/ET/LISA, it will be very hard to distinguish a quantum black hole model from a classical black hole in alternative GR theories or a black hole mimicker.

\vspace{1cm}
\noindent
{\bf Acknowledgments}
\vspace{0.2cm}

I am grateful to Leor Barack, Steve Giddings and Bert Vercnocke for interesting exchanges. I gratefully thank Roberto Emparan for his comments on the tunnelling argument and I also thank an anonymous referee for his/her constructive comments. I thank all the organizers of the conference ``Athens 2019: Gravitational Waves, Black Holes and Fundamental Physics'' for the invitation to present the review talk that led to this article. This work was supported in part by the European Research Council Starting Grant 335146 ``HoloBHC", the convention IISN 4.4503.15. and  the COST Action GWverse CA16104. G.C. is a Research Associate of the Fonds de la Recherche Scientifique F.R.S.-FNRS (Belgium).
%

\vspace{0.5cm}
\noindent
{\bf Bibliography}
\vspace{0.2cm}


\providecommand{\newblock}{}

\end{document}